\documentclass[jkps,preprint, fleqn,showpacs,showkeys]{revtex4}
\usepackage{graphicx}
\usepackage{amssymb}
\usepackage{amsmath}
\usepackage{bm}

\begin{document}
\setcounter{page}{0}
\title[]{The Proton Therapy Nozzles at Samsung Medical Center:\\ A Monte Carlo Simulation Study using TOPAS}
\author{Kwangzoo \surname{Chung}}
\author{Jinsung \surname{Kim}}
\author{Dae-Hyun \surname{Kim}}
\author{Sunghwan \surname{Ahn}}
\author{Youngyih \surname{Han}}
\email{youngyih@skku.edu}
\affiliation{Department of Radiation Oncology, Samsung Medical Center, Sungkyunkwan University, School of Medicine, Seoul 135-710}

\date[]{Received \today}

\begin{abstract}
To expedite the commissioning process of the proton therapy system at Samsung Medical Center (SMC), we have developed a Monte Carlo simulation model of the proton therapy nozzles by using TOol for PArticle Simulation (TOPAS).
At SMC proton therapy center, we have two gantry rooms with different types of nozzles: a multi-purpose nozzle and a dedicated scanning nozzle.   
Each nozzle has been modeled in detail following the geometry information provided by the manufacturer, Sumitomo Heavy Industries, Ltd.
For this purpose, the novel features of TOPAS, such as the time feature or the ridge filter class, have been used, 
and the appropriate physics models for proton nozzle simulation have been defined.
Dosimetric properties, like percent depth dose curve, spread-out Bragg peak (SOBP), and beam spot size, 
have been simulated and verified against measured beam data. 
Beyond the Monte Carlo nozzle modeling, 
we have developed an interface between TOPAS and the treatment planning system (TPS), RayStation.
An exported radiotherapy (RT) plan from the TPS is interpreted by using an interface and is then translated into the TOPAS input text. 
The developed Monte Carlo nozzle model can be used to estimate the non-beam performance, such as the neutron background, of the nozzles.  
Furthermore, the nozzle model can be used to study the mechanical optimization of the design of the nozzle. 
\end{abstract}

\pacs{87.56.-v, 87.53-j, 87.18.Bb}

\keywords{Monte Carlo simulation, Proton therapy, Passive scattering, Pencil-beam scanning, TOPAS}

\maketitle
\thispagestyle{empty}

\section{INTRODUCTION}
The ultimate goal of radiation therapy is to deliver a high dose to the tumor 
while minimizing the dose to the surrounding healthy tissue.
Since the first insight into using fast protons in radiotherapy by R. Wilson in 1946~\cite{RobertWilson1946}, 
the physical characteristics of the Bragg curve have been well studied because of the distinct dosimetric advantages of protons: 
reduced integral dose and improved target volume coverage. 
The reduced integral dose can be achieved due to the fact that there is no exit dose, and 
the highly conformal target volume coverage can be achieved by improved geometric control of the distal fall-off. 
At the same time, the sharp dose fall-off demands a higher accuracy; 
thus, accurate patient set-up, imaging, dose calculation and quality assurance are required in proton therapy.

In general, Monte Carlo dose calculations are more accurate than analytical dose computations. 
Therefore, in proton therapy, the Monte Carlo methods have become important for achieving the most accurate dose calculation. 
Monte Carlo methods can also be used in the study of the particle fluence. 
Verifications of the analytical dose calculation models, the estimates of the relative biological effectiveness (RBE) in patients, 
estimates of the neutron dose, and calculations of neutron shielding are typical Monte Carlo applications in proton therapy.

At the Proton Therapy Center at Samsung Medical Center (SMC), 
two different types of nozzles have been installed: the multi-purpose nozzle and the dedicated scanning nozzle.
The multi-purpose nozzle can deliver proton beams either in a passive scattering mode or a pencil-beam scanning mode. 
The other treatment nozzle is dedicated only to pencil-beam scanning, with an extended vacuum pipe downstream of the nozzle. 
The commissioning of the treatment nozzles will demand numerous sets of measurements. 
Especially for the wobbling mode, there are various options with different combinations of nozzle elements 
to generate spread-out Bragg peaks (SOBPs). 
Thus, the measurements will consume extensive resources. 

Although measurements are the solid basis of commissioning, Monte Carlo simulations can play an important role.  
The main purpose of this project is to build Monte Carlo models of the treatment nozzles to expedite the commissioning process, 
which requires substantial amount of resources, and to support technical developments and clinical operation. 


\section{Experiments and discussion}

\subsection{Monte Carlo Simulation}
We have used TOPAS to build Monte Carlo models of the treatment nozzles.
TOPAS, which stands for TOol for PArticle Simulation~\cite{TOPAS}, 
is a Geant4 application mainly developed for the Monte Carlo simulation of particle therapy nozzles, including proton therapy nozzles.
Geant4 is a framework for simulating the fundamental physical processes at play during the passage of particles through matter~\cite{GEANT4}. 
It is a well-proven toolkit and, it is not only flexible but also includes a complete range of functionalities, 
like tracking, geometry, physics models and hits.
Especially, the provided physics processes cover a comprehensive range. 
Furthermore, it is the result of a worldwide collaboration of physicists and software engineers. 
TOPAS basically inherits all these merits of Geant4 with an extended features like a time feature~\cite{TimeFeature} and a user-friendly interface. 
We have used g4em-standard\_opt3 with high-precision hadronic process physics models to include the low-energy neutron contribution more precisely.

\subsection{Treatment Nozzles at SMC}
\begin{figure}[htb]
\includegraphics[width=16.0cm]{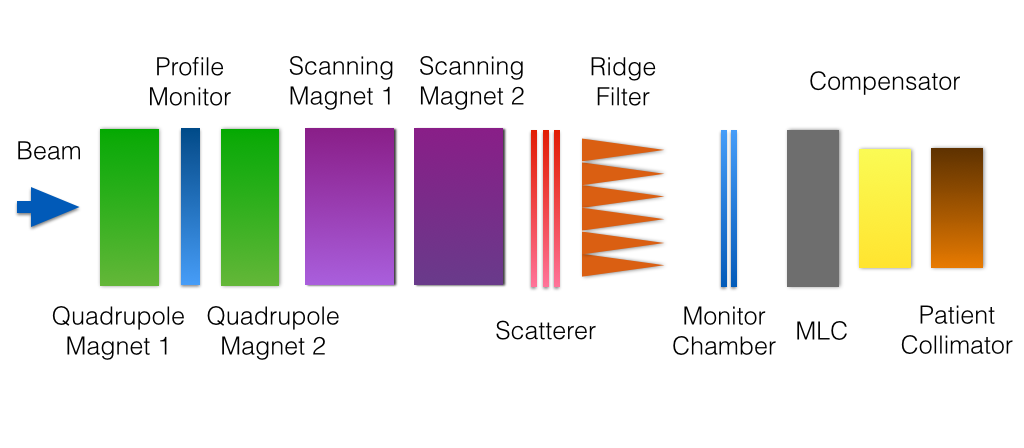}
\includegraphics[width=16.0cm]{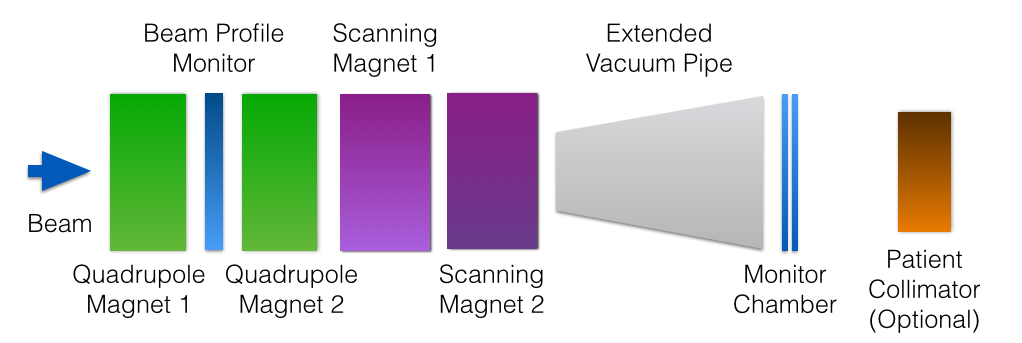}
\caption{(Color online)  
Diagram of the proton therapy nozzles at SMC; the multi-purpose nozzle (top) and the dedicated scanning nozzle (bottom).  
From upstream, there are a pair of quadrupole magnets for focusing and a pair of scanning magnets for wobbling or scanning. 
Note that the beam modulation components of the multi-purpose nozzle have been replaced by an extended vacuum pipe in the dedicated scanning nozzle.
}\label{fig1}
\end{figure}

 Two different types of nozzles have been installed at SMC: 
a multi-purpose nozzle (MPN) and a dedicated pencil-beam scanning nozzle (PBS).
Both nozzles share most of the upstream nozzle elements, such as the quadrupole magnets, scanning/wobbling magnets, 
and beam profile monitors, as shown in Fig.~\ref{fig1}. 

Downstream, the MPN has its own elements for passive scattering of proton beams: 
scatterers, ridge filters, and range compensators.
On the other hand, the PBS has an extended vacuum pipe to preserve the emittance of the proton beams.

The MPN has two different operation modes: a wobbling mode and a scanning mode.
 In the wobbling mode, the two dipole magnets are used with a $90^\circ$ phase shift 
to make a circular trajectory (wobbling) on the scatterers, the trajectory being perpendicular to the direction of the beam's propagation. 
Next, the proton beams pass through the ridge filter after passing the scatterers in order to generate a SOBP. 
Finally, the multi-leaf collimator (MLC), or the aperture, and the compensator shape the lateral and the distal edges of the proton beams.
For the scanning mode, the two dipole magnets are used to control the proton beams in the x and the y directions, and 
scatterers and ridge filters are withdrawn from the proton beam's path. 
One of the special feature in the MPN is the MLC. 
Even though the MLC is an essential part of modern photon radiotherapy, 
only a few proton centers have a MLC for their proton therapy nozzles. 
In the MPN at SMC, the MLC is made of brass, and the leaf thickness of the MLC was designed to be thick enough to stop a 230-MeV proton beam.  
PBS nozzle is dedicated to the pencil-beam scanning mode and provides a larger radiation field size (up to 30 cm x 40 cm).
Downstream, an extended vacuum pipe is placed to maintain a sharp penumbra by suppressing in-air-scattering of proton beam.
Both in the MPN and the PBS, the scanning beam is delivered via continuous line scanning 
with an optional use of patient collimators.

\subsection{Simulation of the Nozzles}

\begin{figure}
\includegraphics[width=7.5cm]{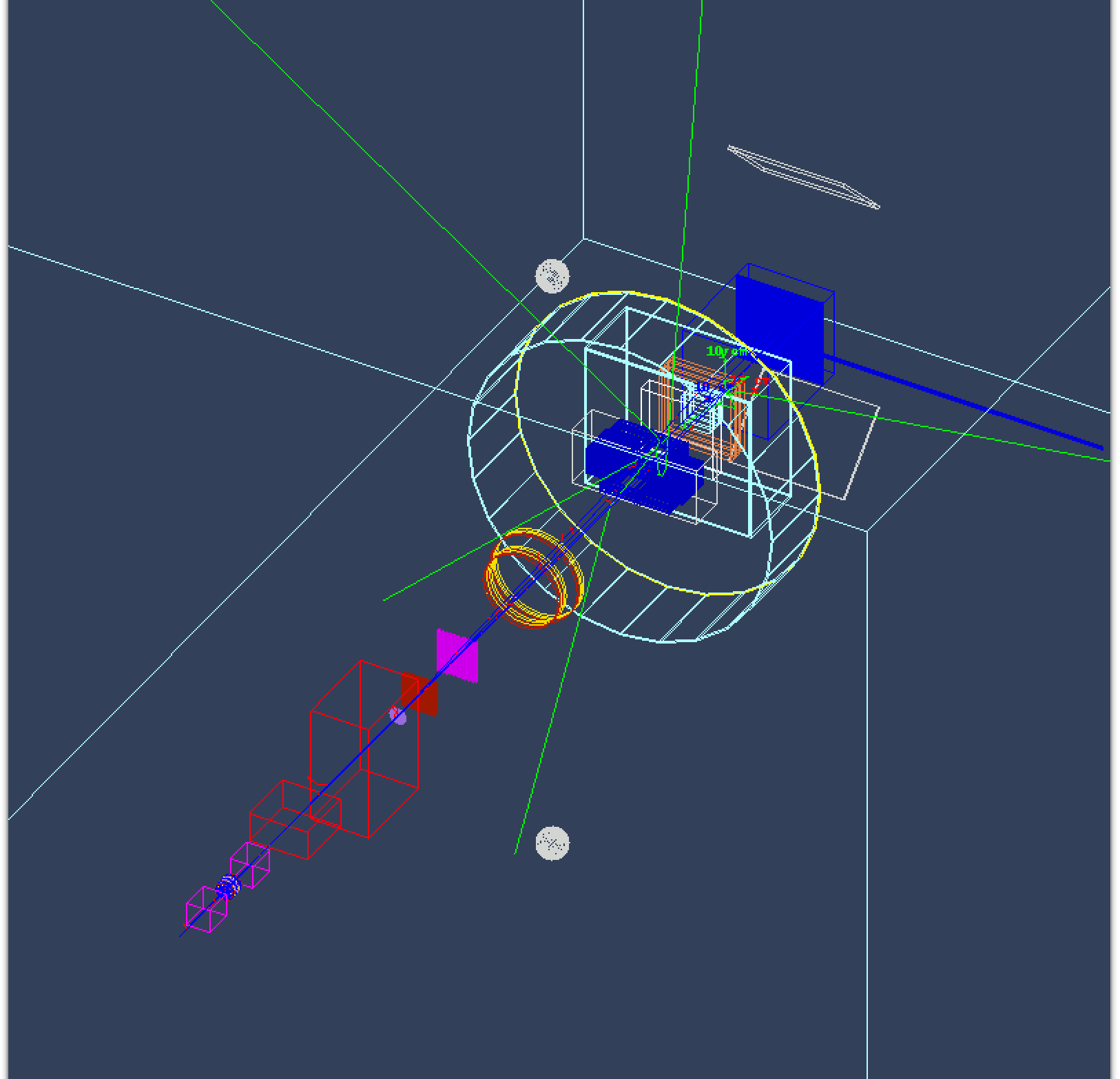}
\hspace{5ex}
\includegraphics[width=7.5cm]{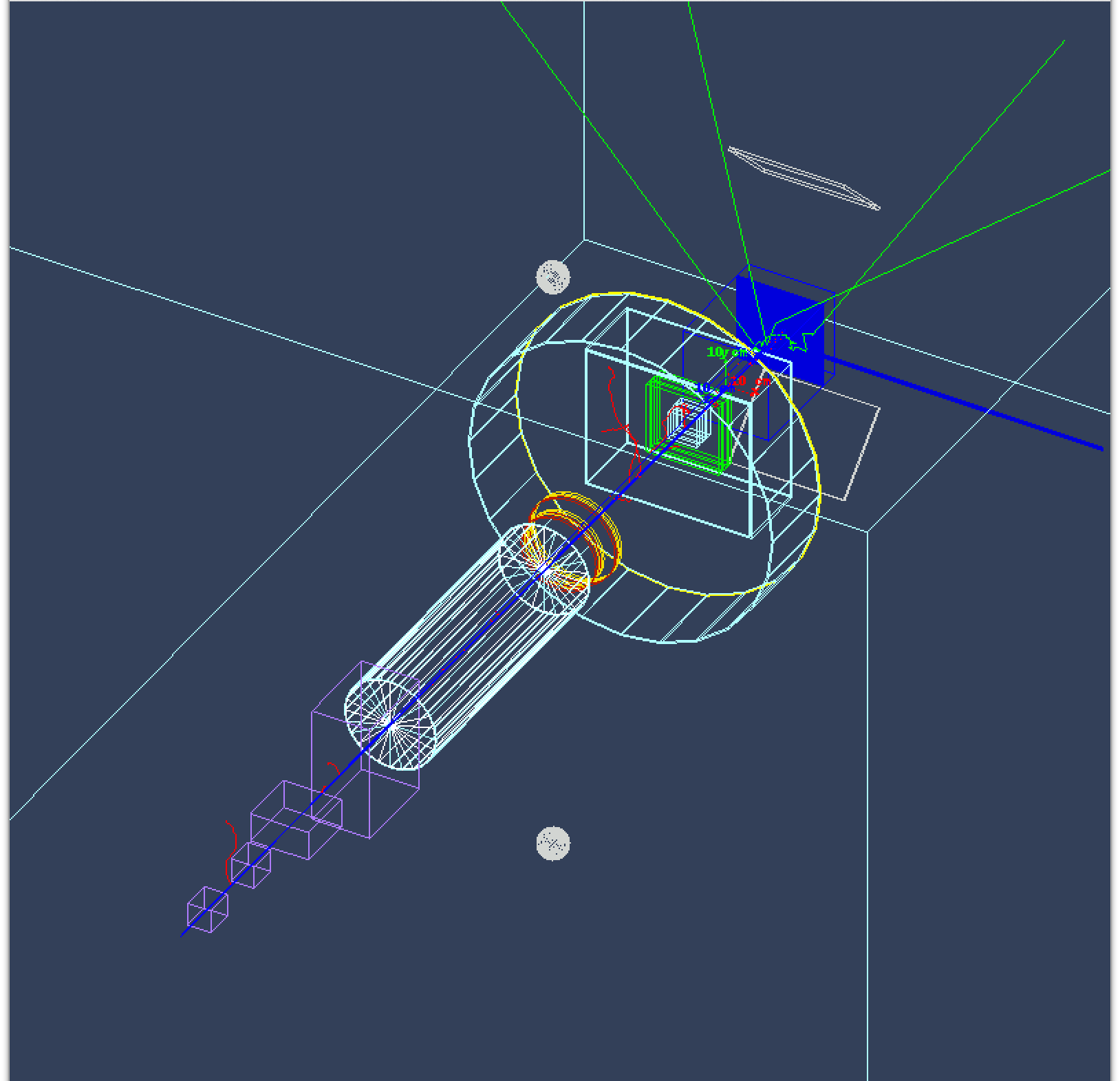}
\caption{(Color online)  
OpenGL visualization of the TOPAS model of the proton therapy nozzles at SMC: 
the multi-purpose nozzle (left) and the dedicated scanning nozzle (right).  
}\label{fig2}
\end{figure}

\begin{figure}
\includegraphics[width=15.0cm]{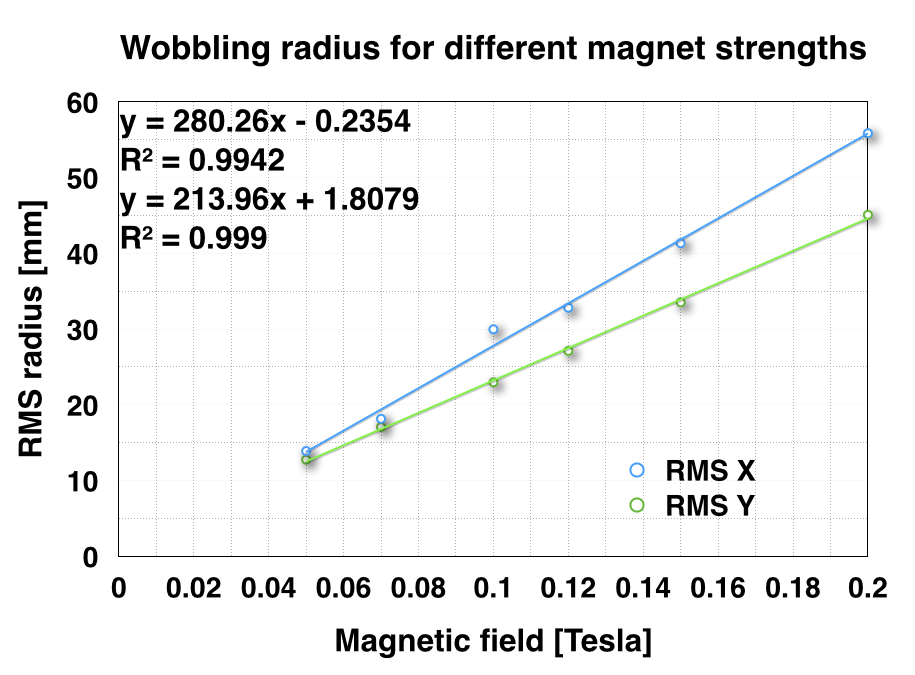}
\caption{(Color online)  
Relation between the magnetic field strength of the dipole magnets and the displacement in the plane of the isocenter.
}\label{fig3}
\end{figure}

Each nozzle element in both nozzles was modeled with sub-millimeter accuracy 
following the detailed information from Sumitomo Heavy Industries, Ltd., as shown in Fig.~\ref{fig2}. 
The MC modules of the common proton nozzle elements, such as the quadrupole magnets and the dipole magnets, 
have been provided by TOPAS.
For the site-specific elements, such as ridge filters and MLC, 
SMC has developed corresponding module classes and has made a contribution to TOPAS. 
For the simulation of wobbling motion, the magnetic field strength of the two dipole magnets (wobbling magnets) 
is varied during the simulation by using the time feature of TOPAS. 
The magnetic field strength of each dipole is varied, and the deviation of the center of the proton beam 
in the plane of the isocenter is determined. 
A linear fit result is used to simulate an arbitrary wobbling radius, as shown in Fig.~\ref{fig3}. 

The scatterer thickness and the wobbling radius need to be determined to generate a uniform proton radiation field. 
The beam size is also a parameter for determining the wobbling radius.
Using TOPAS, we could make possible combinations of parameters with a reasonably uniform radiation field. 
The total scatterer thickness is varied with a resolution of 0.1 mm by using a combination of 7 scatterers. 
The beam spot size for different thicknesses of the scatterer is simulated.

\begin{figure}
\includegraphics[width=8.0cm]{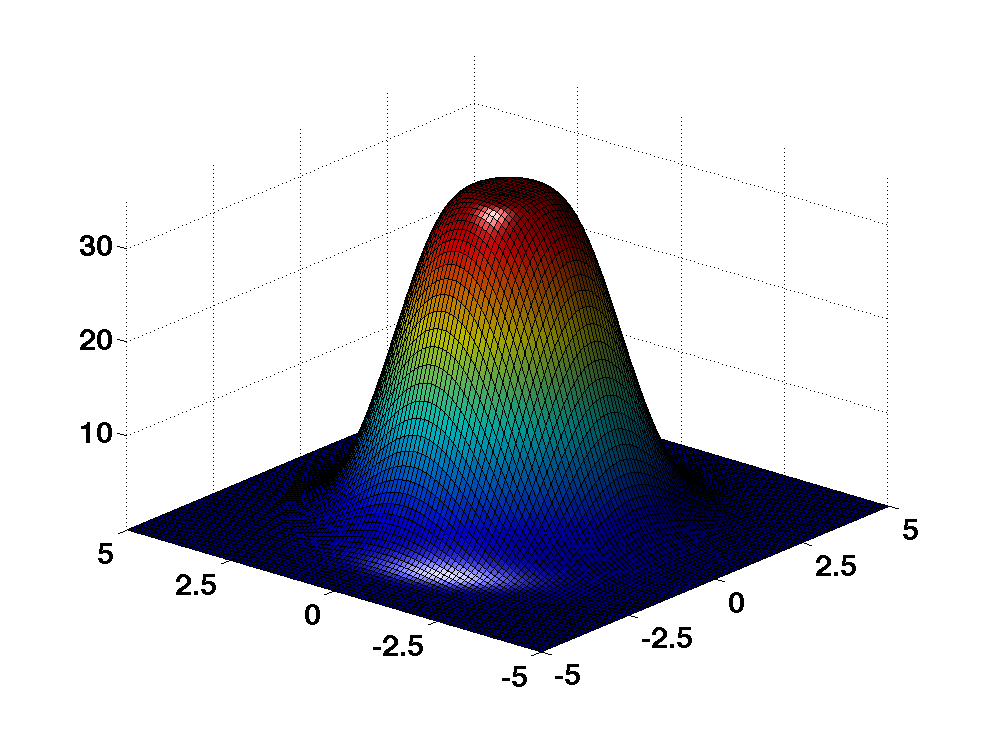}
\includegraphics[width=8.0cm]{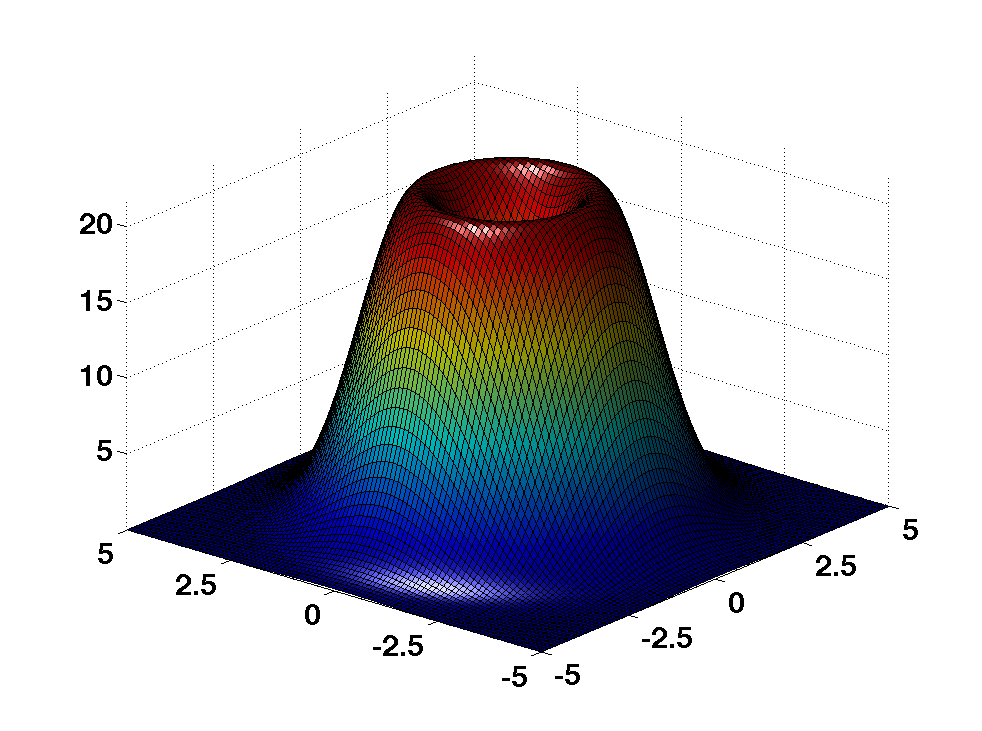}
\caption{(Color online)  
The flat region around the center (left) is used to make a flat dose distribution. 
When the wobbling radius is to large (right), the dose distribution around the center is distorted.   
}\label{fig4}
\end{figure}

The simulated spot size is used to determined the wobbling radius, as shown in Fig.~\ref{fig4}.
If a single Gaussian distribution is assumed, the wobbling radius can be easily estimated analytically.
A medium is known to have a non-negligible contribution from long-range scattering protons; 
thus, a single Gaussian function might fail to describe the exact dose profiles in certain cases~\cite{BeyondGaussians}. 
In our study, a single Gaussian function was enough for the purpose of determining the radii. 
When the wobbling radius is optimized, a possible flat central region will limit the maximum field size. 

Even though a thicker scatterer will provide a larger flat region, 
at the same time, the proton beam will lose more energy and, thus, end up with a shorter range. 
Therefore, trade-off between proton range and field size always exists. 
The optimal combination of these parameters can be decided by using Monte Carlo simulations. 
We actually performed the basic analysis by using TOPAS as mentioned above. 

\begin{figure}
\includegraphics[width=15.0cm]{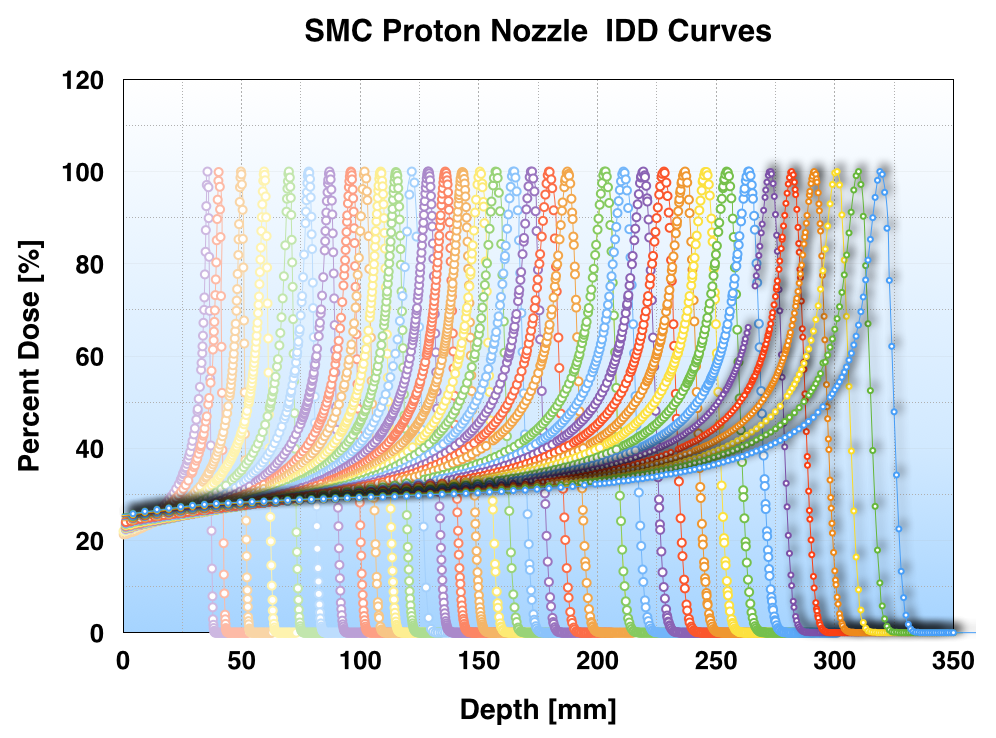}
\includegraphics[width=15.0cm]{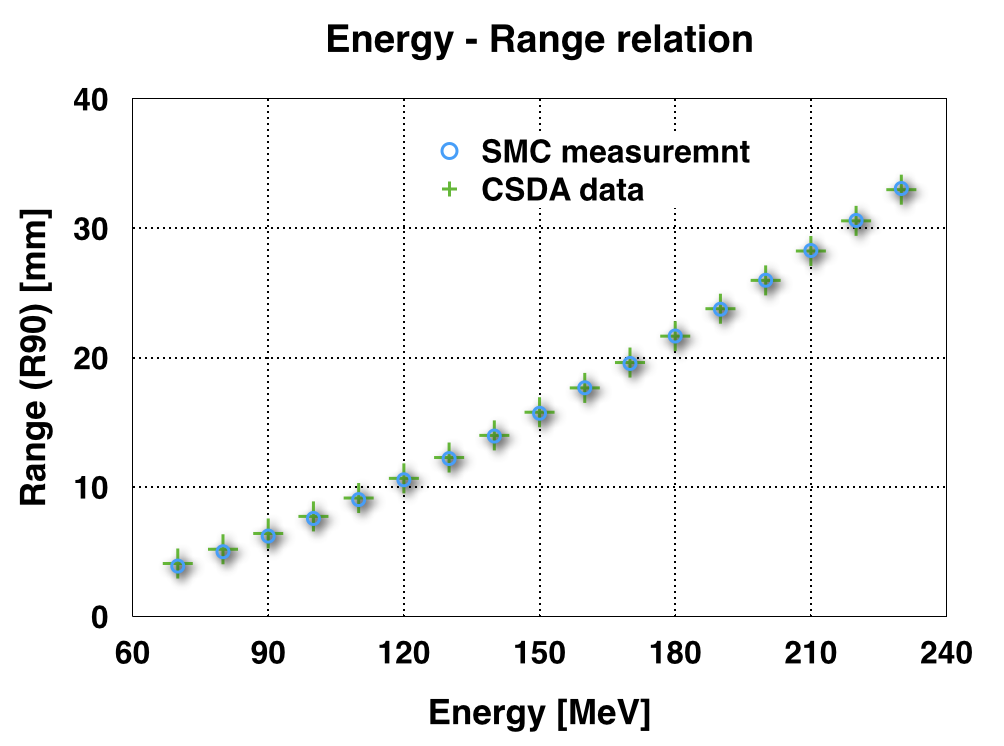}
\caption{(Color online)  
Measured integrated depth dose (IDD) curve of MPN (top) and range comparison with CSDA data (bottom).
}\label{fig5}
\end{figure}

As shown in Fig.~\ref{fig5}, the range-energy relations from the measured integrated depth dose curve of the MPN have been 
compared with the continuous slowing-down approximation (CSDA) of the range~\cite{CSDARange}. 
They agree well and the data are almost on top of each other except in the low-energy region. 
From the difference between the CSDA range and the measured range for a given nominal proton-beam energy, 
the water equivalent thickness (WET) of the scanning nozzle can be decided. 

\subsection{RayStation-TOPAS Interface}
At SMC, RayStation (RaySearch Medical Laboratories AB, Stockholm, Sweden) 
is chosen as the treatment planning system (TPS) for proton therapy.
RayStation provides a Python-based scripting function, 
in which users can not only simplify and customize the planning workflow but also directly interact with core algorithms.
This opens an possibility of exporting the plan parameters in the DICOM~\cite{DICOM} RT plan in the TPS 
and translating them into TOPAS input text.
Ideally, one click of a button on the TPS screen will generate Monte Carlo simulation data using the plan parameters and patient CT images.
At this moment, we have developed a Matlab GUI application which reads in DICOM RT plan files and interprets plan parameters, 
such as the proton beam's energy, gantry�s angle, spot's position, etc.; 
then, the application generates a TOPAS input text file based on the plan parameters.
We are currently working on a Python GUI version of the RayStation-TOPAS interface in the RayStation Python scripting environment. 
We believe that successful implementation will introduce a one-click Monte Carlo simulation era.

\section{CONCLUSIONS}
We have modeled and simulated the two proton therapy nozzles at SMC by using TOPAS. 
Each nozzle elements has been modeled in detail, and various combinations of elements have been studied 
by examining dosimetric properties, such as the integrated depth dose curves and dose profiles.
Using the modeled proton therapy nozzles, we will expand our study not only to dosimetric properties 
but also to design improvement, e.g., ridge filters, MLCs, etc. 
We will perform a validation with the measured data and then use the MC simulation to interpolate/extrapolate the measured data. 
We believe the commissioning process of the proton therapy nozzles at SMC will be expedited by the use of MC simulations. 
Furthermore, the RayStaiton-TOPAS interface will be a valuable tool to validate clinical cases via Monte Carlo simulations.

\section*{ACKNOWLEDGMENTS}
We would like to thank Sumitomo Heavy Industries, Ltd., for providing the detailed geometry of the proton therapy nozzles to be used in the MC simulation, and J. Perl and J. Shin for the support with the TOPAS toolkit. 
This research was supported by the Basic Science Research Program through the National Research Foundation of Korea (NRF), funded by Ministry of Education, Science, and Technology (NRF-2010-0024314) 
and by the National Research Foundation of Korea (NRF) funded by the Ministry of Science, ICT \& Future Planning (2013M2A2A7043507).

\end{document}